\documentclass[12pt, preprint]{aastex}

\slugcomment{submitted to ApJ}

\shorttitle{LARGEST [O III] BLUESHIFTS IN NARROW-LINE QUASARS}
\shortauthors{Aoki, Kawaguchi, \& Ohta}

\begin{document}

\title{The Largest Blueshifts of [\ion{O}{3}] emission line in Two Narrow-Line Quasars}

\author{Kentaro Aoki\altaffilmark{1}}
\affil{Subaru Telescope, National Astronomical Observatory of Japan,
    650 North A'ohoku Place, Hilo, HI 96720}
\email{kaoki@naoj.org}

\author{Toshihiro Kawaguchi\altaffilmark{1,2}}
\affil{LUTH, Observatoire de Paris, Section de Meudon, 5 Place J. Janssen, 92195 Meudon, France}
\email{kawaguti@optik.mtk.nao.ac.jp}

\and

\author{Kouji Ohta\altaffilmark{1}}
\affil{Department of Astronomy, Kyoto University,
Kyoto 606-8502, Japan}
\email{ohta@kusastro.kyoto-u.ac.jp}

\altaffiltext{1}{Visiting Astronomer, Kitt Peak National Observatory, National Optical Astronomy Observatory, which is operated by the Association of Universities for Research in Astronomy, Inc. (AURA) under cooperative agreement with the National Science Foundation.}
\altaffiltext{2}{Postdoctoral Fellow of the Japan Society for the Promotion of Science.}
\begin{abstract}
We have obtained optical intermediate resolution spectra ($R$ = 3000) of the 
narrow-line quasars DMS 0059$-$0055
and PG 1543+489.
The [\ion{O}{3}] emission line in DMS 0059$-$0055 is blueshifted by
880 km s$^{-1}$ relative to H$\beta$.
We also confirm that the [\ion{O}{3}] emission line 
in PG 1543+489 has a relative blueshift of 1150 km s$^{-1}$.
These two narrow-line quasars show the largest [\ion{O}{3}] blueshifts known 
to date among type 1 active galactic nuclei (AGNs). 
The [\ion{O}{3}] emission lines in both objects are broad (1000 -- 2000 km s$^{-1}$) and those in DMS 0059$-$0055 show strong blue asymmetry.
We interpret the large blueshift and the profile of the [\ion{O}{3}] lines as
the result of an outflow
interacting with circumnuclear gas.
Among type 1 AGNs with large blueshifted [\ion{O}{3}], there is no correlation 
between the Eddington ratios and the amount of [\ion{O}{3}] blueshifts.
Combining our new data with published results,
we confirm that the Eddington ratios of the such AGNs are the highest among AGNs 
with the same black hole masses.
These facts suggest that the Eddington ratio is a necessary condition 
or the [\ion{O}{3}] blueshifts weakly depend on the Eddington ratio.
Our new sample suggests that there are possible necessary conditions to produce an outflow
besides a high Eddington ratio: large black hole mass ($ > 10^{7} M_{\sun}$)
or high mass accretion rate ($> 2 M_{\sun}$/yr) 
or large luminosity ($\lambda L_{\lambda} (5100 {\rm \AA}) > 10^{44.6}$ erg s$^{-1}$). 
\end{abstract}


\section{INTRODUCTION}
Narrow-Line Seyfert 1 galaxies (NLS1s) are a subclass of Active
Galactic Nuclei (AGNs) which have characteristics as follows
\citep[see][]{p00}. 
1.) They have relatively narrower permitted lines
(FWHM of H$\beta \le$ 2000 km s$^{-1}$) than those of 
``normal'' Broad-Line Seyfert 1 (BLS1s).
The 2000 km s$^{-1}$ in FWHM of H$\beta$ is usually used to separate 
NLS1s and BLS1s.
2.) They often emit strong optical \ion{Fe}{2}
multiplets or higher ionization iron lines, that are seen in Seyfert 1s 
but not seen in Seyfert 2s \citep{OP85}.
3.) Their X-ray spectra are significantly softer (photon indices in soft X-ray are 
$\Gamma = 1.5 - 5$) than those of BLS1s ($\Gamma \sim 2.1$) \citep{Bol96, Lao97}.
4.) They show rapid soft/hard X-ray variability \citep{lei99a}.
The most likely interpretation of the characteristics is that 
for a given luminosity NLS1s contain less massive black holes with high accretion rates 
than BLS1s \citep{Bo02}.
NLS1s tend to have larger Eddington ratios, which is the ratio of its 
bolometric luminosity to Eddington luminosity, than BLS1s \citep[e.g.][]{kaw03}.
\par
Outflow phenomena have been reported in some NLS1s and narrow-line quasars (NLQs)
which are luminous ($M_B < -23$ magnitudes) counterparts of NLS1s.
I Zw 1, the prototype of NLS1s/NLQs, has been known to have
[\ion{O}{3}] $\lambda\lambda4959, 5007$ emission lines blueshifted by 600 km s$^{-1}$
relative to other emission lines
\citep{Phi76}.
The redshift of H$\beta$ in I Zw 1 is consistent with the redshift of 
the centroid of \ion{H}{1} 21 cm line \citep{cond85} and the molecular CO line \citep{bar89},
that represent the systemic redshift of the object more reliably.
The [\ion{O}{3}] emission line in I Zw 1 is thus blueshifted relative to not only H$\beta$ 
but also the systemic velocity.
\citet{zam02} discovered that six other AGNs show [\ion{O}{3}] blueshifts relative to H$\beta$ 
that are larger than 250 km s$^{-1}$ among a sample of 216 type 1 AGNs.
We hereafter refer to them as ``blue outliers'' following \citet{zam02}.
The broad H$\beta$ lines of ``blue outliers''are narrow ($\leqq 4000$ km s$^{-1}$), and a half of them are NLQs.
The large [\ion{O}{3}] blueshift is interpreted as the result of the outflow whose
receding part is obscured by an optically thick accretion disk \citep{zam02}.
The origin of the outflow is considered to be a strong radiation pressure due to 
a large Eddington ratio \citep{gru02, zam02}.
\par
Although more than 10 ``blue outliers'' have been found 
\citep{gru02, zam02, mar03a},
there is no detailed observations of their [\ion{O}{3}] line profiles.
Higher resolution data are crucial to identify their weak [\ion{O}{3}] emission lines
which are blended with strong \ion{Fe}{2} emission lines.
The previous observations were carried out with a resolution of 5 \AA~ -- 7 \AA,
i.e., resolving power of $R=1000$ -- $1500$.
Therefore, we have carried out higher resolution spectroscopy 
($R = 3000$) of two NLQs,
DMS 0059$-$0055 and PG 1543+48 around the H$\beta$ -- [\ion{O}{3}]$\lambda5007$ region.
These two NLQs have the largest known [\ion{O}{3}] blueshifts among type 1 AGNs.
It is important to observe the objects with the most extreme blueshift
to understand the [\ion{O}{3}] outflow phenomena. 
If the origin of the outflow is the radiation pressure,
objects with more luminous and/or higher Eddington ratios are expected to show larger blueshifts.
\citet{mar03a} indeed showed that ``blue outliers'' have higher Eddington ratios among 
AGNs with the similar black hole masses.
\par
DMS 0059$-$0055 ($z=0.295$) was discovered by 
the Deep Multicolor CCD Survey \citep{ken97}.
DMS 0059$-$0055 is included as SDSS J010226.31-003904.6 in the NLS1s sample of \citet{Wil02}.
They searched the Sloan Digital Sky Survey \citep[SDSS;][]{Yor00} 
Early Data Release \citep[EDR;][]{Sto02} for NLS1s, and compiled a sample of 150 NLS1s.
The FWHM of H$\beta$ of DMS 0059$-$0055 is 1680 km s$^{-1}$ \citep{Wil02}.
Its spectrum also shows prominent \ion{Fe}{2} emission lines.
Its soft X-ray spectrum is steep with a photon index of 3.2 \citep{Wil02}.
These characteristics of DMS 0059$-$0055 are typical for NLS1s/NLQs.
\par
PG 1543+489 ($z=0.401$) is a radio-quiet quasar in the Bright Quasar Survey \citep{sch83}.
It shows a narrow H$\beta$ (FWHM=1600 km s$^{-1}$) emission and strong \ion{Fe}{2} emissions
\citep{BG92}.
Its soft X-ray emission has a steep spectral slope 
with a photon index of 3.1 \citep{Lao97}.
Therefore PG 1543+489 also has typical characteristics of NLS1s/NLQs.
It has the largest [\ion{O}{3}] blueshift (950km s$^{-1}$) relative to H$\beta$ 
among the 280 type 1 AGNs in the sample of \citet{mar03a}.
The [\ion{O}{3}] is so weak that the [\ion{O}{3}] emission line analyzed by 
\citet{zam02} has a large uncertainty.
This is the reason why we thought a confirmation with a higher S/N ratio and dispersion was necessary.
\par
Here we report the results for spectroscopy of DMS 0059$-$0055 and PG 1543+489.
In \S~\ref{Obs} our KPNO observations and the spectrum of DMS 0059$-$0055 in the SDSS EDR
are described.
The emission line properties are presented in \S~\ref{Results}.
In \S~\ref{Discussion} we discuss the correlations with 
the [\ion{O}{3}] blueshift of the Eddington ratio and the optical luminosity 
among ``blue outliers''.
We also compare the Eddington ratios and the black hole masses of
AGNs with and without the [\ion{O}{3}] blueshift, 
and discuss the possible conditions to produce [\ion{O}{3}] outflow.
We give concluding remarks in \S~\ref{Conclusion}.
We assume $H_{\rm 0} = 70$ km s$^{-1}$ Mpc$^{-1}$, 
$\Omega_{m} = 0.3$, and $\Omega_{\Lambda} = 0.7$ throughout this paper.

\section{OBSERVATIONS AND DATA REDUCTION} \label{Obs}
\subsection{KPNO Data}
The spectra of DMS 0059$-$0055 and PG 1543+489 were obtained with the GoldCam spectrograph 
of the KPNO 2.1-m telescope on 2002 September 16 (UT) and 2003 June 18 (UT), respectively.
The nights were under photometric condition and the seeing was $\sim 2$\arcsec.
We binned the F3KC CCD along the slit direction, 
so the projected pixel size was 1\farcs56 along the slit.
The \#56 grating (600 lines mm$^{-1}$) was used in the second order and 
the sampling was 0.65 \AA~ pixel$^{-1}$.
A slit width of 2\farcs0 was used, resulting in a resolution of 2.0 \AA~
measured by night sky lines ($R\sim3000$).
The observed spectral range was 6000 -- 7250 \AA,
which covers 
from \ion{Fe}{2} $\lambda4667$ to \ion{Fe}{2} $\lambda5534$ and
from \ion{Fe}{2} $\lambda4385$ to \ion{Fe}{2} $\lambda5018$ 
in the rest-frame spectra of DMS 0059$-$0055 and PG 1543+489, 
respectively.
The slit position angle was 90\arcdeg.
We obtained $5 \times 1800$ s and $6 \times 1800$ s exposures 
for DMS 0059$-$0055 and PG 1543+489, respectively.
A spectrophotometric standard star Wolf 1346 or HD 109995 was observed 
for flux calibration at the beginning of the night.
\par
The data were reduced using IRAF\footnote{IRAF is distributed by the National Optical Astronomy
Observatories, which is operated by the Association of Universities for Research in Astronomy, Inc. (AURA) under
cooperative agreement with the National Science Foundation.}
in the standard manner for CCD data.
After subtracting the bias level, 
each spectrum frame was flat fielded with the internal quartz lamp frames,
and then bad columns and cosmic-ray events were removed.
Wavelength calibration of each object frame was performed using the comparison lamp frames.
A small wavelength offset was corrected by using night sky lines.
The rms wavelength calibration error is 0.05 \AA~and 
it corresponds to 3 km s$^{-1}$ at 6300 \AA~ where H$\beta$ and [\ion{O}{3}] 
are redshifted to in these objects.
After sky subtraction, flux calibration was performed.
The sensitivity along wavelength was calibrated within an uncertainty of 3 -- 4 \%. 
One dimensional spectra were extracted, and all five and six spectra were combined to 
one spectrum each for DMS 0059$-$0055 and PG 1543+489.
The atmospheric B band absorption overlapped with 
\ion{Fe}{2} $\lambda5317$ of DMS 0059$-$0055 
and \ion{Fe}{2} $\lambda4924$ of PG 1543+495.
The absorption feature was removed using the standard star's spectra.
The Galactic interstellar reddening was corrected using $E(\bv) = 0.036$ mag. and 0.018 mag. \citep{Sch98} 
for DMS 0059$-$0055 and PG 1543+489, respectively, and 
the empirical selective extinction function by \citet{Car89} 
and $R_{V} \equiv A(V)/E(\bv) = 3.1$.

\subsection{SDSS Data}
The spectrum of DMS 0059$-$0055 in the SDSS EDR was obtained on 2000 September 29 
with the 2.5m telescope at Apach Point Observatory.
The spectral resolution is $\sim 1800$ and the wavelength coverage is 3800 -- 9200 \AA~
\citep{Sto02}.
The flux and wavelength calibration was done by the spectroscopic pipelines \citep{Sto02}.
The interstellar reddening was corrected using $E(\bv) = 0.036$ mag. which is the same value 
used for the KPNO data.

\subsection{\ion{Fe}{2} Subtraction}
Both NLQs show strong \ion{Fe}{2} emission lines that
must be removed for the measurement of [\ion{O}{3}]
emission line.
The \ion{Fe}{2} emission lines were subtracted by the template method by \citet{BG92}.
The \ion{Fe}{2} template was made from a spectrum of I Zw 1.
We observed I Zw 1 with the same configuration as PG 1543+489 and DMS 0059$-$0055
except for the wavelength coverage ($\lambda\lambda$ 4200 -- 5400 \AA).
The power-law continuum was determined at 4200 \AA~ and 4750 \AA~ 
where \ion{Fe}{2} contamination is relatively weak.
After subtracting the continuum, H$\beta$, H$\gamma$, [\ion{O}{3}] $\lambda\lambda 4959, 5007$,
and [\ion{Fe}{2}] emission lines were fitted and subtracted.
Balmer lines were fitted with a Lorentzian plus a Gaussian, and [\ion{Fe}{2}] emission lines were fitted 
with a Lorentzian.
The [\ion{O}{3}] emission lines of I Zw 1 are known to be broad and cannot be fitted with 
a single Gaussian \citep{VVG01}.
We thus fitted the [\ion{O}{3}] $\lambda\lambda 4959, 5007$ using two sets of two Gaussians. 
The width and redshift were fixed for each set, and the intensity ratio of 
[\ion{O}{3}] $\lambda 5007$ to $\lambda 4959$ was fixed to be 3.0.
\par
The power-law continuum was subtracted from the spectrum of PG 1543+489 and DMS 0059$-$0055
by using the same method applied to I Zw 1.
Next, the \ion{Fe}{2} template was fitted to, and subtracted from 
the continuum subtracted spectrum.
The \ion{Fe}{2} template was fitted by changing scaling and broadening factors
within reasonable ranges at the red half of the \ion{Fe}{2} $\lambda 5018$ 
emission line where the contamination of H$\beta$ and [\ion{O}{3}] is the smallest.
\par
\section{RESULTS} \label{Results}
\subsection{DMS 0059$-$0055}
\begin{figure}
\epsscale{0.8}
\plotone{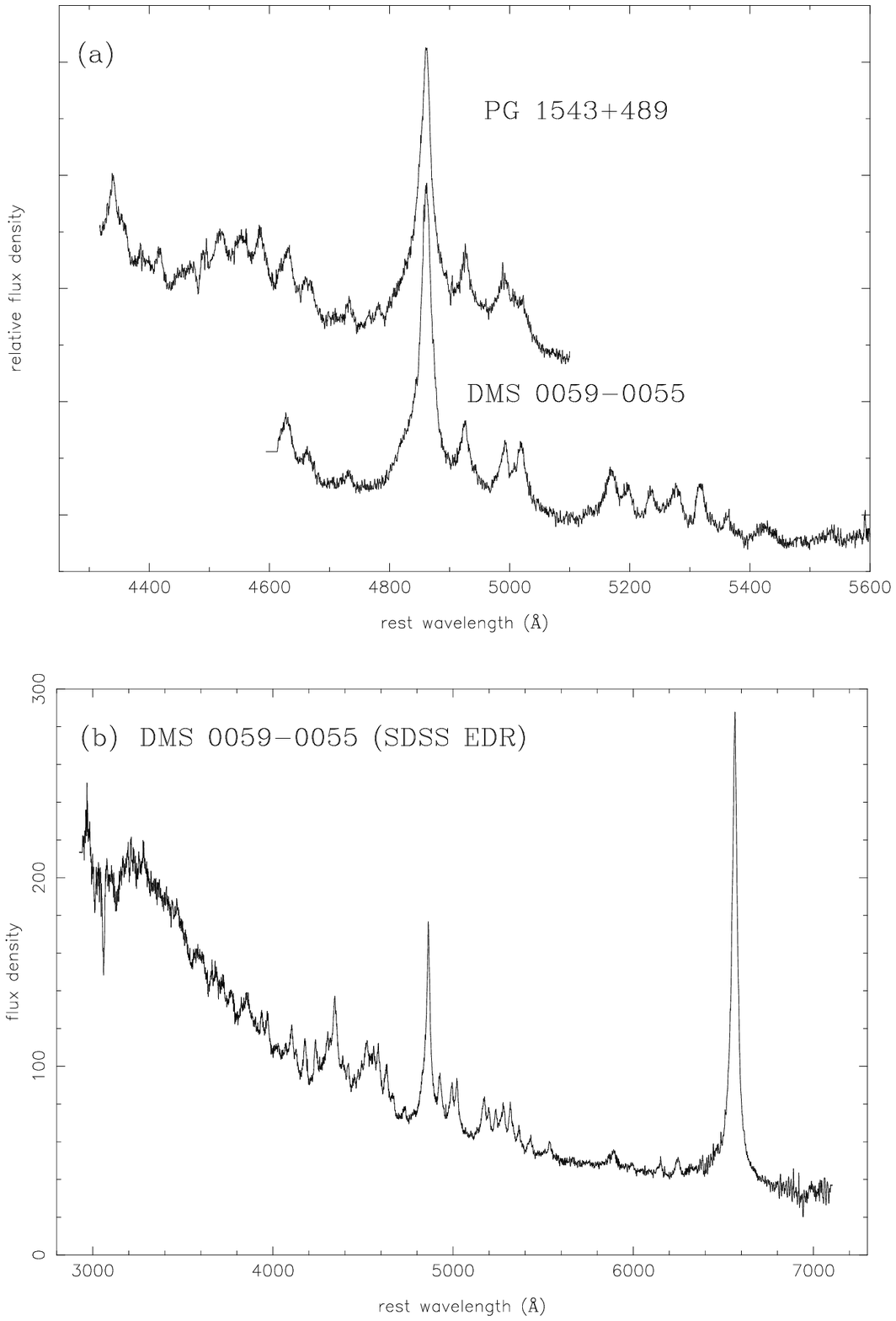}
\caption{The deredshifted spectra of DMS 0059$-$0055 and PG 1543+489.
(a) The spectra of DMS 0059$-$0055 and PG 1543+489 obtained with KPNO 2.1m.
Ordinate is a relative flux density in units of erg s$^{-1}$ cm$^{-2}$ \AA$^{-1}$,
and abscissa is a rest wavelength in angstrom.
Spectra are normalized and shifted in arbitrary values.
(b) The spectrum of DMS 0059$-$0055 in the SDSS EDR. 
Ordinate is a flux density in units of $10^{-17}$ erg s$^{-1}$ cm$^{-2}$ \AA$^{-1}$
and abscissa is a rest wavelength in angstrom.\label{fig1}}
\end{figure}

\begin{figure}
\plotone{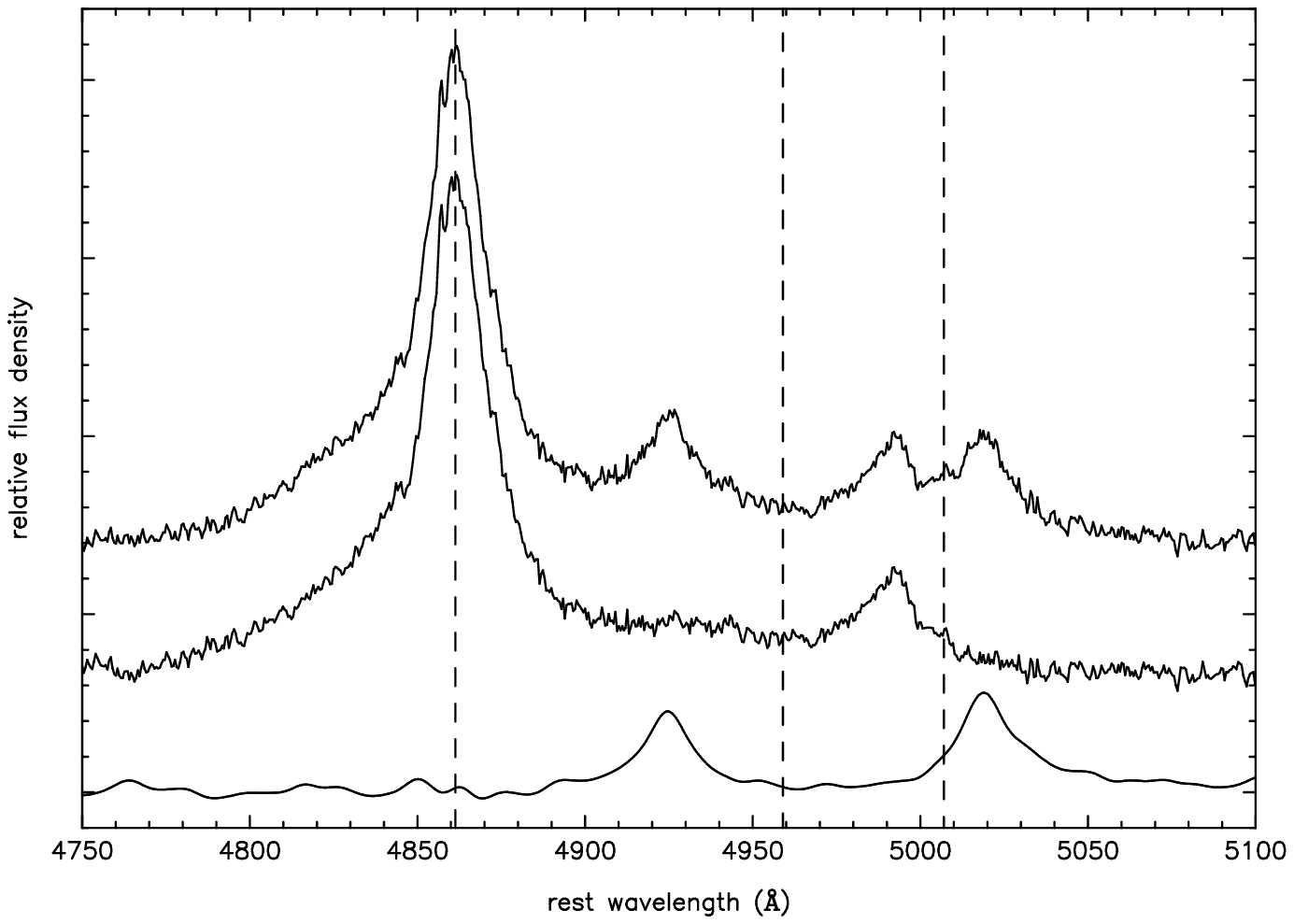}
\caption{\ion{Fe}{2} subtracted spectrum of DMS 0059$-$0055. 
Ordinate is a relative flux density in units of erg s$^{-1}$ cm$^{-2}$ \AA$^{-1}$
and abscissa is a rest wavelength in angstrom.
The upper most is the continuum subtracted spectrum,
the middle is the \ion{Fe}{2} subtracted one, and the lower is
the \ion{Fe}{2} template used.
These three spectra are shifted by adding arbitrary constants.
The dashed lines mark the wavelengths of H$\beta$ and [\ion{O}{3}] $\lambda\lambda
4959, 5007$ expected from the redshift of H$\beta$ peak.
\label{fig2}}
\end{figure}

\begin{figure}
\plotone{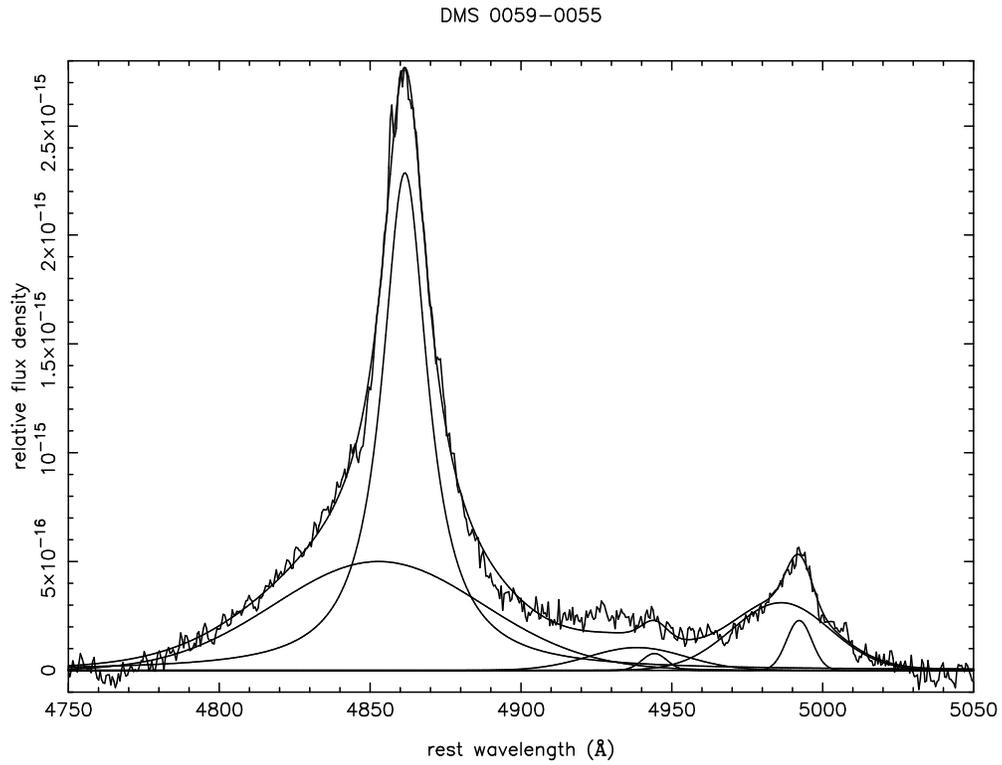}
\caption{The fitting in the H$\beta$ -- [\ion{O}{3}] region of DMS 0059$-$0055. 
Ordinate is a relative flux density in units of erg s$^{-1}$ cm$^{-2}$ \AA$^{-1}$
and abscissa is a rest wavelength in angstrom.
H$\beta$ is fitted with a Lorentzian and a Gaussian, and
[\ion{O}{3}] doublet is fitted with two sets of two Gaussians.
Note that the continuum was subtracted.
\label{fig3}}
\end{figure}

The rest--frame spectra of DMS 0059$-$0055 obtained with KPNO 2.1-m and SDSS EDR are 
shown in Figure \ref{fig1}.
The redshift for H$\beta$ emission was used to the deredshift.
Figure \ref{fig2} displays the \ion{Fe}{2} subtracted spectrum
around the H$\beta$ -- [\ion{O}{3}] region (middle) obtained with KPNO 2.1-m 
as well as the \ion{Fe}{2} template (bottom) and 
the continuum subtracted spectrum (top) before the \ion{Fe}{2} template subtraction.
In the middle spectrum there is an emission line whose peak is at $4992.2\pm1.0$ \AA~ 
and the width is comparable to those of \ion{Fe}{2} and H$\beta$.
If this feature is the blueshifted [\ion{O}{3}] $\lambda5007$ emission line
as found in several NLQs \citep{zam02, mar03a},
the corresponding [\ion{O}{3}] $\lambda 4959$ should be seen at 4944.2 \AA~
with one-third of the flux of the $\lambda4992$ feature.
The [\ion{O}{3}] $\lambda 4959$ at first glance does not seem to exist in 
the \ion{Fe}{2} subtracted spectrum.
However, we can obtain an acceptable fit to the spectrum with [\ion{O}{3}] $\lambda\lambda 4959, 5007$
together with H$\beta$ as shown in Figure \ref{fig3}.
We thus identify the $\lambda4992$ feature with the [\ion{O}{3}] $\lambda5007$ emission line.
Since the $\lambda4992$ feature has an asymmetric profile 
(i.e. not well fitted with a single Gaussian), 
[\ion{O}{3}] $\lambda\lambda 4959, 5007$ were fitted with two 
Gaussians for each line.
H$\beta$ was fitted with a combination of a Lorentzian and a Gaussian.
The FWHMs are corrected for the instrumental broadening using the simple assumption:
$\rm{FWHM}_{true}=(\rm{FWHM}_{obs}^{2} - \rm{FWHM}_{inst}^{2})^{1/2}$,
where $\rm{FWHM}_{obs}$ is an observed FWHM of a line and $\rm{FWHM}_{inst}$ is an 
instrumental FWHM.
The fitting result is shown in Figure \ref{fig3}.
One [\ion{O}{3}] component with $\rm{FWHM} = 2300\pm150$ km s$^{-1}$ is blueshifted by 
$1240\pm100$ km s$^{-1}$ relative to H$\beta$, and 
the other component with $\rm{FWHM} = 580\pm70$ km s$^{-1}$ is blueshifted by $880\pm30$ km s$^{-1}$.
The broad profile and large blueshift merge [\ion{O}{3}] $\lambda 4959$ into 
the red wing of H$\beta$ broad line and smear [\ion{O}{3}] $\lambda 4959$. 
The large blueshifted [\ion{O}{3}] emission line in DMS 0059$-$0055 has been newly discovered 
in this work.
The [\ion{O}{3}] $\lambda5007$ profile shows asymmetry and has strong blue tail.
The directly measured FWHM of [\ion{O}{3}] $\lambda5007$ is $1320\pm80$ km s$^{-1}$ while that of H$\beta$ is $1500\pm50$ km s$^{-1}$.
\par
The \ion{Fe}{2} subtracted spectrum from the SDSS EDR was consistent with the KPNO spectrum
except for the line width of the narrower component of [\ion{O}{3}].
One [\ion{O}{3}] component with $\rm{FWHM} = 2310\pm130$ km s$^{-1}$ is blueshifted by 
$1200\pm80$ km s$^{-1}$ (relative to H$\beta$), and 
the other component with $\rm{FWHM} = 840\pm150$ km s$^{-1}$ is blueshifted by $860\pm40$ km s$^{-1}$.
The spectral resolution of the SDSS EDR spectrum is factor $\sim 2$ lower than
that of our KPNO spectrum
while the S/N of the SDSS EDR spectrum is comparable to that of our KPNO spectrum.
The line width of the narrower component of [\ion{O}{3}] in the SDSS spectrum is significantly 
broader than that in the KPNO spectrum.
Although we could not find any reason for the discrepancy in the line width,
we prefer the result from the KPNO spectrum because it was obtained 
with a higher spectral resolution and a comparable S/N.
Since the SDSS spectrum was absolute flux calibrated, we can measure the
[\ion{O}{3}] flux.
The [\ion{O}{3}]$\lambda5007$ flux is $(1.4\pm0.5) \times 10^{-15}$ erg s$^{-1}$ cm$^{-2}$ of the narrower component, 
and $(5.7\pm0.5) \times 10^{-15}$ erg s$^{-1}$ cm$^{-2}$ of the broader one.
\par
In order to estimate the ratio of [\ion{O}{3}] $\lambda5007$/H$\beta$ 
for the [\ion{O}{3}] emitting gas,
we tried to fit the spectrum with the components described above plus two Gaussians 
which represent blueshifted H$\beta$ 
from the [\ion{O}{3}] emitting gas.
The FWHM and redshift were fixed to be the same values of the corresponding [\ion{O}{3}] component.
The intensity ratio of the broad component and the narrow one was also fixed to the same
value as [\ion{O}{3}] component.
However, the fitting result did not converge.
We could not estimate the flux of the blueshifted H$\beta$ corresponding to 
the [\ion{O}{3}] component.
\par
In order to measure \ion{Fe}{2} emission line between 4434 -- 4684 \AA
~(hereafter \ion{Fe}{2} $\lambda4570$),
we used the continuum subtracted SDSS spectrum.
The flux of \ion{Fe}{2} $\lambda4570$ is $4.4 \times 10^{-14}$ erg s$^{-1}$ cm$^{-2}$.
Since the flux of H$\beta$ is $(4.3\pm1.1) \times 10^{-14}$ erg s$^{-1}$ cm$^{-2}$,
the flux ratio of \ion{Fe}{2} $\lambda4570$/H$\beta$ is 1.0.
The large \ion{Fe}{2} $\lambda4570$/H$\beta$ makes DMS 0059$-$0055 a strong \ion{Fe}{2} emitter.
This fact is consistent with the indication by \citet{zam02} that ``blue outliers''
tend to show strong \ion{Fe}{2} emission.
\subsection{PG 1543+489}
\begin{figure}
\plotone{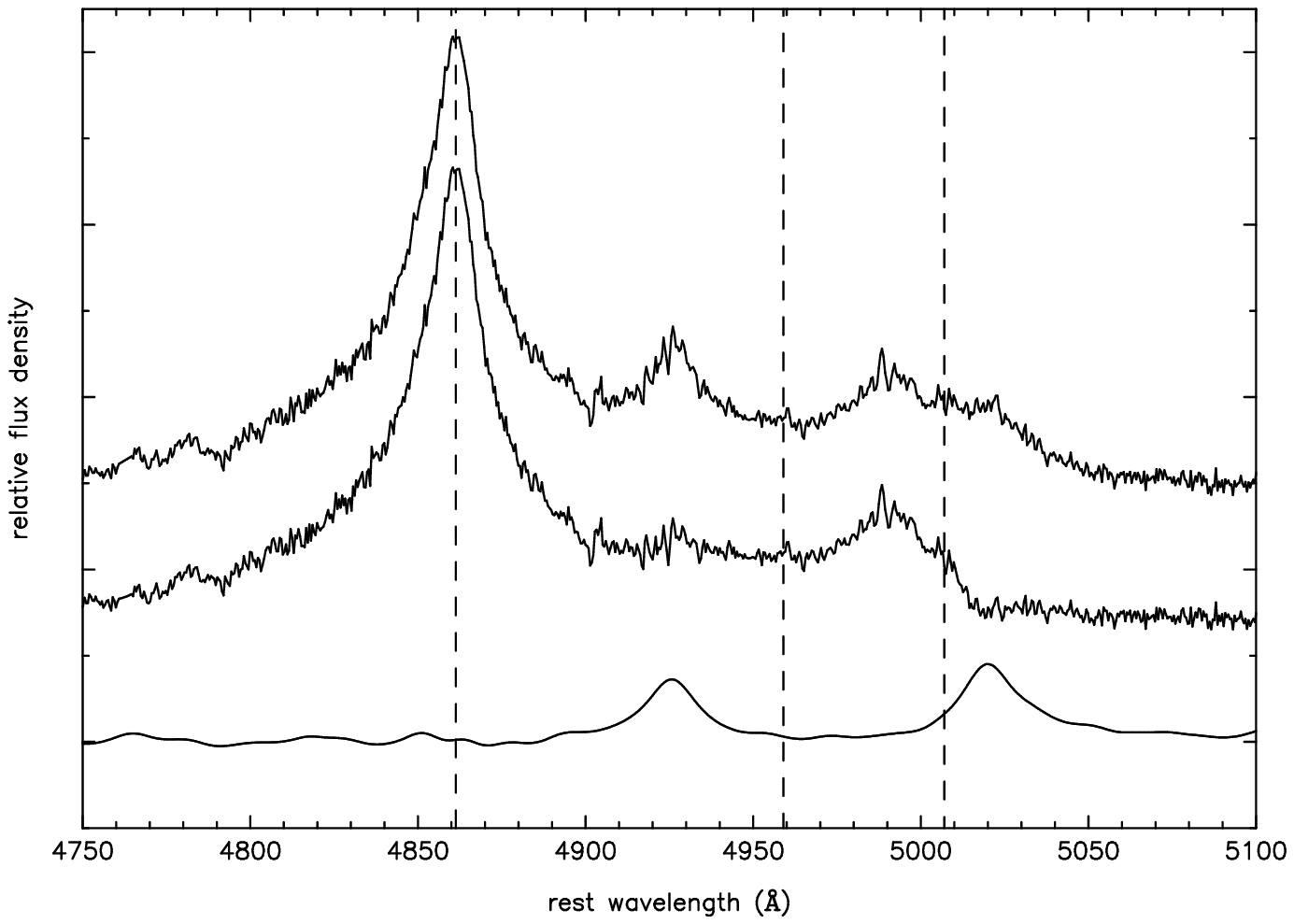}
\caption{\ion{Fe}{2} subtracted spectrum of PG 1543+489. 
Ordinate is a relative flux density in units of erg s$^{-1}$ cm$^{-2}$ \AA$^{-1}$
and abscissa is a rest wavelength in angstrom.
The upper most is the continuum subtracted spectrum,
the middle is the \ion{Fe}{2} subtracted one, and the lower is
the \ion{Fe}{2} template used.
These three spectra are shifted by adding arbitrary constants.
The dashed lines mark the the wavelengths of H$\beta$ and [\ion{O}{3}] $\lambda\lambda
4959, 5007$ expected from the redshift of H$\beta$ peak.
\label{fig4}}
\end{figure}

\begin{figure}
\plotone{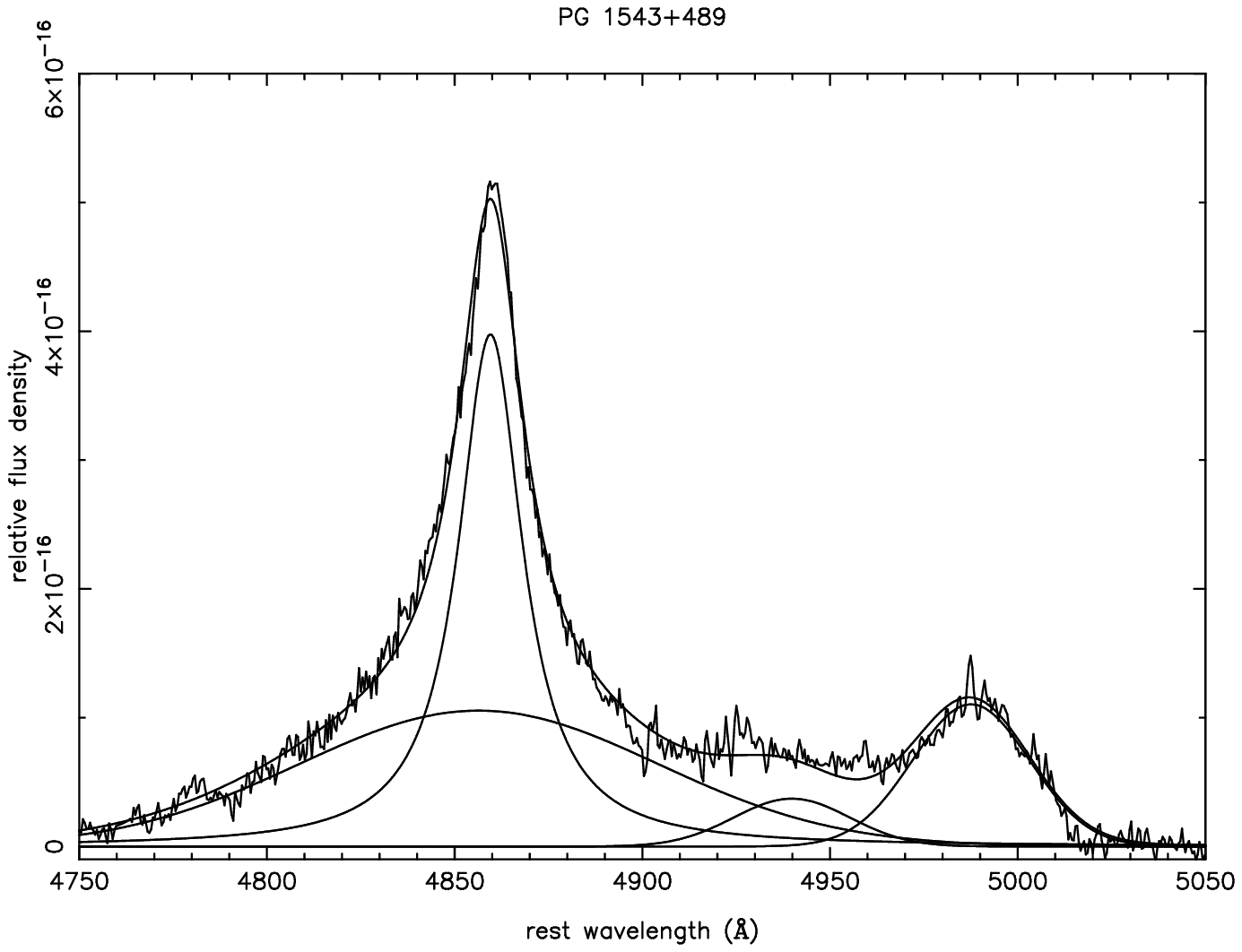}
\caption{The fitting in the H$\beta$ -- [\ion{O}{3}] region of PG 1543+489. 
Ordinate is a relative flux density in units of erg s$^{-1}$ cm$^{-2}$ \AA$^{-1}$
and abscissa is a rest wavelength in angstrom.
H$\beta$ is fitted with a Lorentzian and a Gaussian, and
[\ion{O}{3}] doublet is fitted a Gaussian.
Note that the continuum was subtracted.
\label{fig5}}
\end{figure}

The spectrum of PG 1543+489 deredshifted by referring to the H$\beta$ peak is 
shown in Figure \ref{fig1}.
Figure \ref{fig4} displays the \ion{Fe}{2} subtracted spectrum
around the H$\beta$ -- [\ion{O}{3}] region (middle) obtained with KPNO 2.1-m as well as 
the \ion{Fe}{2} template (bottom) used and the continuum subtracted spectrum (middle) 
before the \ion{Fe}{2} template subtraction.
Our spectrum was taken with a three times higher spectral resolution 
than that by \citet{BG92}.
The fitting result for PG 1543+489 around the H$\beta$ -- [\ion{O}{3}] region is shown 
in Figure \ref{fig5}.
The spectrum of PG 1543+489 is similar to that of DMS 0059$-$0055,
except for the absence of a blue tail in [\ion{O}{3}] line in PG 1543+489.
H$\beta$ was fitted with a Lorentzian plus a Gaussian.
[\ion{O}{3}] $\lambda\lambda 4959, 5007$ were fitted with a broad 
($\rm{FWHM} = 2200\pm40$ km s$^{-1}$) blueshifted ($1150 \pm 20$ km s$^{-1}$ relative to H$\beta$) 
Gaussian.
This [\ion{O}{3}] blueshift is 20\% larger than that reported previously \citep{zam02, mar03a}.
Comparing Figure 2 in \citet{zam02} and our Figure \ref{fig5}, our spectrum has a higher S/N ratio 
as well as a three times higher resolution.
Our blueshift measurement thus seems more reliable, and 
the disagreement is probably due to improvement of data quality.
The directly measured FWHM of H$\beta$ is $1630\pm50$ km s$^{-1}$
which is consistent with 1600 km s$^{-1}$ \citep{zam02} and 1560 km s$^{-1}$ 
\citep{BG92, mar03a}.
The absolute flux of [\ion{O}{3}] could not be measured for PG 1543+489
since the spectroscopy at KPNO was done with the comparable slit width to the seeing size.
\par
\begin{figure}
\plotone{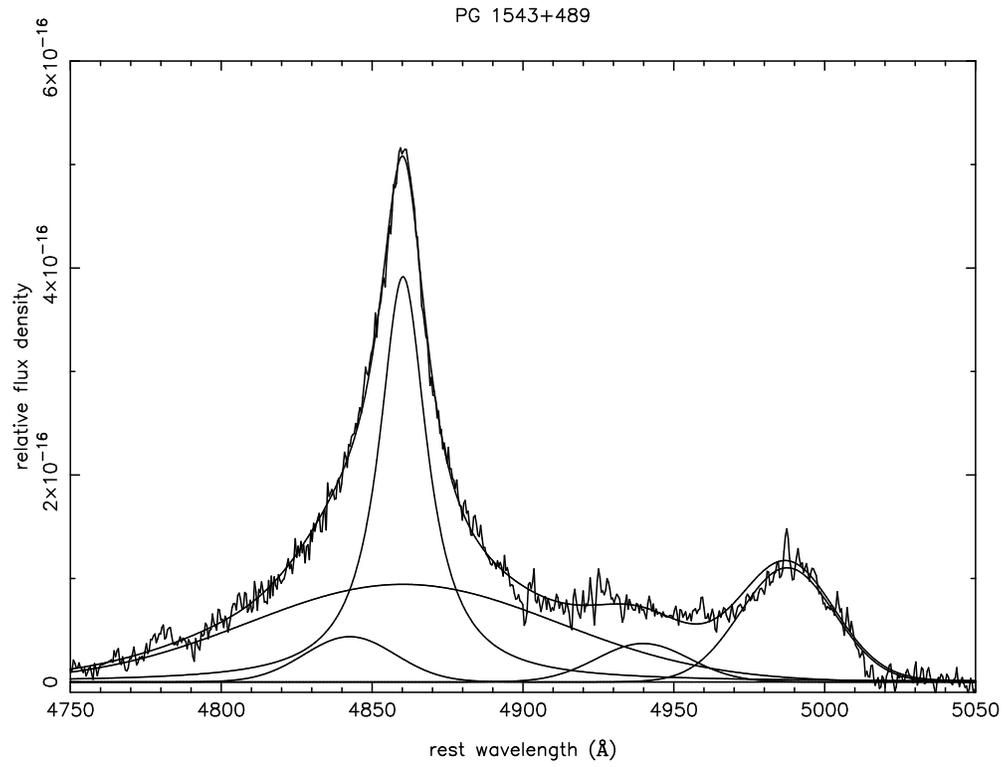}
\caption{Same as Figure 5, but added a H$\beta$
which is blueshifted by the same amount with [\ion{O}{3}] emission.
\label{fig6}}
\end{figure}

As done for DMS 0059$-$0055, we tried to detect a blueshifted H$\beta$ emission line.
Adding a single Gaussian which represents a blueshifted H$\beta$ with the same width and 
redshift as the [\ion{O}{3}],
we have fitted a spectrum as shown in Figure \ref{fig6}.
The [\ion{O}{3}]$\lambda5007$/H$\beta$ is 0.38.
Although the fit converged in this case,
this blueshifted H$\beta$ line is completely buried under the blue wing of H$\beta$.
Therefore we think that the blueshifted H$\beta$ is only marginally detected, 
and that its flux has a large uncertainty.
\par
We measured the flux of \ion{Fe}{2} $\lambda4570$ from the continuum subtracted spectrum.
The flux ratio of \ion{Fe}{2} $\lambda4570$ to H$\beta$ is 0.85, and agrees well 
with \citet{BG92}, and 30\% larger than \citet{zam02}.
Although \citet{BG92} and \citet{zam02} analyzed the same spectrum, 
their \ion{Fe}{2} $\lambda4570$/H$\beta$ are different; 
0.86 \citep{BG92} and 0.64 \citep{zam02}.
Considering such a discrepancy in measurements from the same data,
our result is consistent with the previous measurements.

\section{THE NECESSARY CONDITIONS FOR [\ion{O}{3}] OUTFLOWS} \label{Discussion}
The blueshift velocity of [\ion{O}{3}] in DMS 0059$-$0055 ($880$ km s$^{-1}$) is comparable 
to that of PG 1543+489 ($1150$ km s$^{-1}$) 
which has the largest blueshift of [\ion{O}{3}] among $\sim 280$ type 1 AGNs
\citep{mar03a}.
Our spectroscopy reveals that the [\ion{O}{3}] lines in both 
DMS 0059$-$0055 and PG 1543+489 are broad (1000 -- 2000 km s$^{-1}$).
Especially, a strong blue tail is prominent in the [\ion{O}{3}] line of DMS 0059$-$0055.
The [\ion{O}{3}]
blueshift relative to H$\beta$ together with the strong blue tail suggests that the [\ion{O}{3}] is emitted from
outflowing gas from the nuclei
and the receding part of the flow is obscured
\citep{zam02}.
The large width suggests that the outflow interacts with the surrounding gas.
Although it is difficult to reject completely the possibility that optically thick 
gas is inflowing towards the black hole and 
the inflowing gas at the near side of the black hole is not seen,
we prefer the outflow hypothesis.
If the inflowing gas is Thomson thick ($> 10^{24}$ cm$^{-2}$), 
the redshifted absorption line should be observed in some ``blue outliers''.
If the inflowing gas is dusty and optically thick, nuclear broad lines and a continuum may be significantly
reddened.
However, neither phenomena are observed in ``blue outliers''.
\par
Since DMS 0059$-$0055 and PG 1543+489 have the largest [\ion{O}{3}] blueshifts known,
we would expect them to have extreme characteristics among ``blue outliers''.
These extreme objects would enable us to understand the origin
of the [\ion{O}{3}] outflow.
In order to search for a clue to the origin of the [\ion{O}{3}] outflow,
we compiled ``blue outliers'' from the literature. and have examined correlations between the velocity shift of [\ion{O}{3}] relative to H$\beta$
(hereafter referred to as $\Delta$v)
and several other properties 
(the Eddington ratio, the optical luminosity etc.) among ``blue outliers''.
\citet{mar03a} measured $\Delta$v among the type 1 AGN samples of \citet{mar03b}, 
which contains 215 $z \la 0.8$ AGNs, 
and \citet{gru99}, which is a soft-X ray selected sample.
They found 12 objects with $\Delta{\rm v} < -250$ km s$^{-1}$ and identified them as ``blue outliers''.
Since \citet{zam02} found that all `` blue outliers'' have FWHM H$\beta \la 4000$ km s$^{-1}$,
we looked for ``blue outliers'' among NLS1s/NLQs in \citet{VVG01} and \citet{lei99b}.
We do not think that \citet{zam02} missed ``blue outliers'' in BLS1s/QSOs 
because it is easier to measure the [\ion{O}{3}] emission line in BLS1s/QSOs
than in NLS1s/NLQs;
BLS1s/QSOs have stronger [\ion{O}{3}] emission and weaker \ion{Fe}{2} emission
which contaminates to [\ion{O}{3}] emission than NLS1s/NLQs.
We have found that IRAS 04416+1215 and IRAS 04576+0912 have $\Delta{\rm v} < -250$ km s$^{-1}$
from figure 1 of \citet{VVG01}.
\citet{gru02} reported IRAS 13224$-$3809 has $\Delta{\rm v} = -370$ km s$^{-1}$.
In the end, we compiled a list of 16 ``blue outliers'' including DMS 0059$-$0055 from our new data.
We tabulated $\Delta$v, FWHM of [\ion{O}{3}] $\lambda5007$, FWHM of H$\beta$, 
and their references of the 16 ``blue outliers'' in Table~\ref{tbl1} in the order of their $\Delta$v.
\par
The optical luminosity, $\lambda L_{\lambda}(5100{\rm \AA})$, of ``blue outliers'' was estimated from the $B$ magnitude 
given in \citet{VV03} and the Galactic extinction $A_{B}$ which is calculated from the E(\bv) values
of \citet{Sch98} and given in NED
\footnote{The NASA/IPAC Extragalactic Database (NED) is operated by the Jet Propulsion
Laboratory, California Institute of Technology, under contract with the
National Aeronautics and Space Administration.}.
The $k-$correction was done by assuming $f_{\nu} \propto \nu^{-0.44}$ \citep{VB01}.
The Eddington ratio was calculated from the Eddington luminosity, $1.3 \times 10^{38} M_{BH}/M_{\sun} $
erg s$^{-1}$, and 
the bolometric luminosity, $L_{bol}$, which is assumed to be 
$13 \times \lambda L_{\lambda}$(5100 \AA) erg s$^{-1}$ \citep{Elv94}.
The black hole mass, $M_{BH}$, was estimated using the relation between size of 
the broad-line region and the luminosity and the assumption of virialized motions
in broad-line regions \citep{kas00} as follows:
$$ M_{BH} = 4.817 \times 10^6 \left(\frac{{\rm FWHM~ H}\beta}{10^{3}~ {\rm km~ s}^{-1}}\right)^{2} 
\left(\frac{\lambda L_{\lambda}(5100{\rm \AA})}{10^{44}~ {\rm erg~ s}^{-1}}\right)^{0.7} M_{\sun}. $$
Note that most of NLS1s/NLQs in \citet{kas00} also follow the same relation
as BLS1s/QSOs.
Therefore, black hole mass estimation with H$\beta$ width and optical luminosity
\citep{Wandel99,kas00} is likely to be valid not only for BLS1s/QSOs but also for NLS1s/NLQs.
We simply assume (as is done by \citet{mar03a}) that this black hole mass
estimation can also work for ``blue outliers'', 
although the relation between the size of broad-line region and the luminosity for 
``blue outliers'' have not been obtained observationally. 
The lack of apparent difference in H$\beta$  profiles between ``blue
outliers'' and  other NLS1s/NLQs without [\ion{O}{3}] outflow supports this
assumption.
\par
We also examine their radio properties; 
we took radio flux data from the literature for nine ``blue outliers'', and 
calculated the monochromatic radio luminosity at 5 GHz and the radio loudness
parameter, log $R$.
Here, $R$ is the ratio of radio (5 GHz) to optical ($B$-band) flux density.
Table ~\ref{tbl1} also gives $\lambda L_{\lambda}$(5100 \AA),
$M_{BH}$, the Eddington ratio,
the radio luminosity and the radio loudness parameter.
\par
\begin{figure}
\plotone{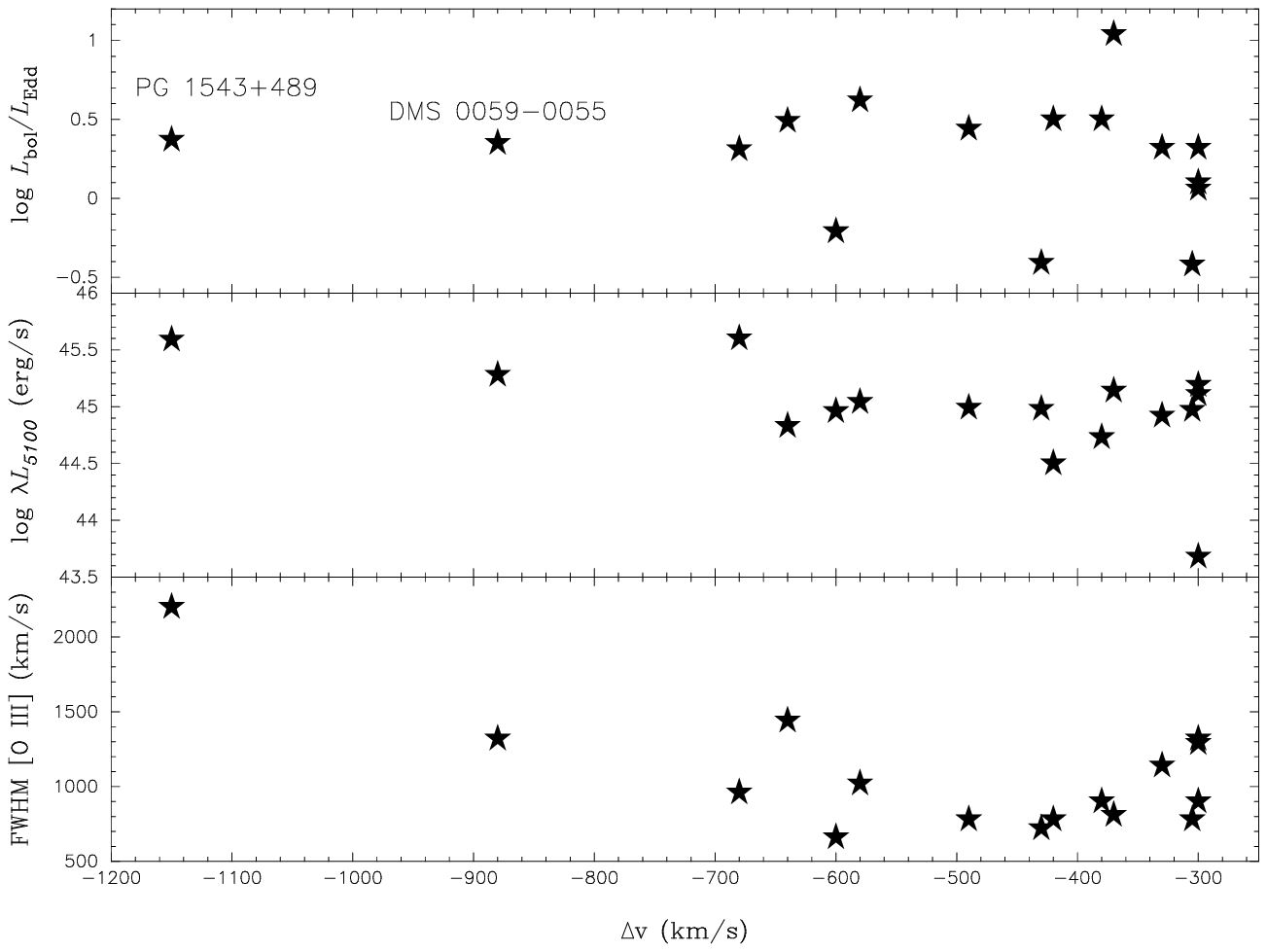}
\caption{The blueshift velocity ($\Delta$v) vs. the Eddington ratio, the optical luminosity and 
[\ion{O}{3}] line width
from top, middle, and low, respectively.
No property does not significantly correlate with $\Delta$v.
\label{fig7}}
\end{figure}

Figure \ref{fig7} shows relations between Eddington ratio, optical luminosity, and
[\ion{O}{3}] line width versus $\Delta$v.
No clear correlation is seen between the Eddington ratio and $\Delta$v (the top panel of
Fig. \ref{fig7}),
rather the ratio is roughly constant for various $\Delta$v.
The optical luminosity (the middle panel of Fig. \ref{fig7}), and the [\ion{O}{3}] line width
(the bottom panel of Fig. \ref{fig7}) seem to correlate with $\Delta$v;
large $\Delta$v tends to occur in large luminosity and large [O III] width.
However, the Spearman rank correlation coefficients, $r_{S}$, and its probabilities from
null correlation, P$_{\rm r}$, for these distributions 
show no significant correlation (Table~\ref{tbl2}).
Since neither the Eddington ratio nor the optical luminosity correlates with $\Delta$v,
there is no strong luminosity dependence of $\Delta$v among ``blue outliers''.
We suspect that the lack of any strong correlations with $\Delta$v may indicate 
that the $\Delta$v is affected by the viewing angle from us \citep{mar03a}.
This may also explain the distribution of data points in the middle 
panel of Fig. \ref{fig7}; as optical luminosity increases, the absolute value 
of the maximum blueshift among ``blue outliers'' of a given luminosity also 
increases.
The maximum blueshift may occur at the smallest viewing angle.
The $\Delta$v does not depend on the radio properties;
neither the radio power nor radio loudness correlates with $\Delta$v.
\par
\begin{figure}
\plotone{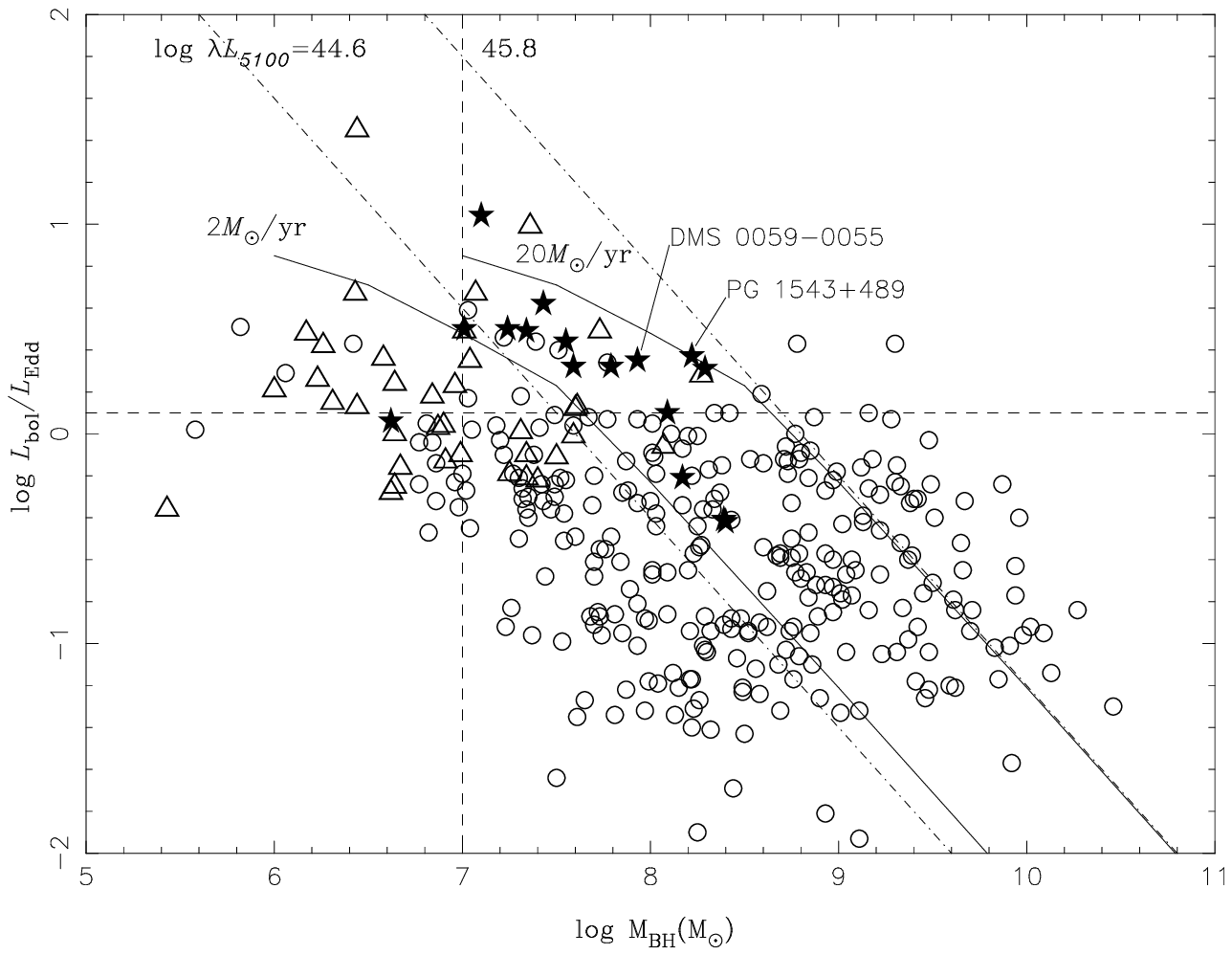}
\caption{Plot of black hole mass vs. Eddington ratio; NLS1s from \citet{VVG01} and \citet{lei99b}
are coded as open triangles, ``blue outliers'' as stars,
and type 1 AGNs from \citet{mar03a} as open circles.
The dashed lines indicate log $L_{bol}/L_{Edd} = 0.1$, and $M_{\rm BH} = 10^{7} M_{\sun}$,
respectively. 
The solid lines show the Eddington ratio for various $M_{\rm BH}$
with a given $\dot{M}$.
The dot-dashed lines represent the constant optical luminosity.
\label{fig8}}
\end{figure}

Although \citet{mar03a} pointed out that
``blue outliers'' have the highest Eddington ratios among their sample,
we could not find any significant correlation between $\Delta$v and the Eddington ratio
among ``blue outliers''.
Moreover, we have found that a number of NLS1s/NLQs in \citet{VVG01} and \citet{lei99b}
with high Eddington ratios are not ``blue outliers''.
In order to search for other possible conditions necessary for the large [\ion{O}{3}] outflow
besides the high Eddington ratio,
we plotted our ``blue outliers'' sample, NLS1s/NLQs in \citet{VVG01} and \citet{lei99b} as well as 
the \citet{mar03a} sample, in the $M_{BH}$ vs. Eddington ratio plane (Fig. \ref{fig8}).
The $M_{BH}$ and Eddington ratio of this sample were calculated in the same manner 
as our ``blue outliers'' sample.
While we find that there are no correlations between $\Delta$v and the Eddington ratio among ``blue outliers'',
we confirmed that the ``blue outliers'' tend to have higher Eddington ratios than AGNs without [O III] outflows.
This fact suggests that either the high Eddington ratio is a necessary condition,
or the dependence of $\Delta$v on the Eddington ratio is weak and may be affected by viewing angle.
\par
Inspecting Figure \ref{fig8} prompts us to consider $M_{BH}$ besides Eddington ratio 
as a condition to produce the [\ion{O}{3}] outflow.
The \citet{mar03a} sample includes only a few objects with a small black hole mass ($< 10^{7} M_{\sun}$) 
and a high Eddington ratio.
Our sample includes more such objects.
A considerable number of objects with $M_{BH} < 10^{7} M_{\sun}$ also have 
high Eddington ratios (log $L_{bol}/L_{Edd} \geq 0.1$).
None of these 15 objects is, however,  a ``blue outlier''.
On the other hand, there are a dozen ``blue outliers'' 
among the 33 objects with $M_{BH} \ge 10^{7} M_{\sun}$ and 
high Eddington ratios.
Thus, it seems that $\sim 10^{7} M_{\sun}$ is the lower limit of $M_{BH}$ for
producing the [\ion{O}{3}] outflow.
\par
The high Eddington ratio and the large $M_{BH}$ seem to be the conditions necessary for
an [\ion{O}{3}] outflow.
The distribution of ``blue outliers'' in Figure \ref{fig8}, however, hints at 
the presence of another condition. 
There is a continuous distribution of ``blue outliers'' from (log $M_{BH}$, log $L_{bol}/L_{Edd}$) = 
(7, 0.5) to (8.4, $-0.4$) in Figure \ref{fig8}. 
This distribution implies
that there is a possible lower limit in the mass accretion rate ($\dot{M}$) or
the optical luminosity to produce the [\ion{O}{3}] outflow.
The Eddington ratio decreases with increasing $M_{BH}$ at a constant $\dot{M}$.
We show the Eddington ratio for various $M_{\rm BH}$ at a given $\dot{M}$ 
with solid lines in  Figure \ref{fig8}.
Bolometric luminosity is computed with the accretion disk model by
\citet{kaw03}, which is applicable to both sub-Eddington and 
super-Eddington ($\dot{M} \gg L_{\rm Edd}/c^2$) accretion rates.
The Eddington ratio is inversely proportional to $M_{\rm BH}$
at high-$M_{\rm BH}$ range (sub-Eddington accretion rate regime),
resulting from a constant bolometric luminosity for a given $\dot{M}$
(i.e. constant radiative efficiency).
The bolometric luminosity saturates at a few times Eddington luminosity at low-$M_{\rm BH}$
range (super-Eddington accretion rate regime), because an accretion disk
becomes radiatively inefficient as $\dot{M} / (L_{\rm Edd}/c^2)$ increases
\citep{Abr88}.
All ``blue outliers'' but one (IRAS 04576+0912)
\footnote{The $B$ magnitude, 16.58, of IRAS 04576+0912 may be incorrect.
The $J$ and $K_{s}$ magnitudes are 13.8 and 11.9, respectively,
from 2MASS Point Source Catalog.
After Galactic reddening correction, $B-K_{s}=4.1$ magnitude while $J-K_{s}=1.9$ magnitude.
$B-K_{s}$ should be $\sim3$ magnitude from $J-K_{s}$ at $z=0.037$ adopting the mean-QSO
spectrum by \citet{Elv94}.
If the luminosity increases by 1 magnitude,
the black hole mass also increases by 0.28 dex.
So the data point moves +0.12 dex along y-axis and +0.28 dex along x-axis
in Figure \ref{fig8}
and would be much closer to the line. 
}
are at the right hand side of the left solid line in Figure \ref{fig8},
which indicates $\dot{M}\sim 2 M_{\sun} /$yr.
Thus, $\dot{M}\sim 2 M_{\sun} /$yr seems to be the lower limit on the mass accretion rate necessary for producing a large
[\ion{O}{3}] outflow.
\par
The slopes of the lines representing constant $\dot{M}$ and 
constant optical luminosity on the $M_{BH}$ vs. the Eddington ratio plane are
so similar that it is difficult to distinguish how these two variables affect 
the presence of ``blue outliers''.
The Eddington ratio also decreases when $M_{BH}$ increases at a constant optical 
luminosity
since the Eddington ratio is expressed as follows:
$${\rm log} L_{bol}/L_{Edd} = -{\rm log} M_{BH} (M_{\sun}) + {\rm log} \lambda L_{\lambda} (5100 {\rm \AA}) ({\rm erg~ s}^{-1}) -37.$$
The dot-dashed lines represent constant optical luminosities in Figure \ref{fig8}.
Most ``blue outliers'' are located at the right hand side of the constant 
log $\lambda L_{\lambda} (5100 {\rm \AA}) ({\rm erg~ s}^{-1}) = 44.6$,
indicating that 
log $\lambda L_{\lambda} (5100 {\rm \AA}) ({\rm erg~ s}^{-1}) \ga 44.6$ is also necessary
for the large [\ion{O}{3}] outflow. 
We conclude that not only Eddington ratio but also $M_{BH}$ or $\dot{M}$ or luminosity
may control the [\ion{O}{3}] outflow.
However other factors (e.g., viewing angle) may govern $\Delta$v because among ``blue outliers''
there is no correlation between
$\Delta$v and the Eddington ratio or the optical luminosity.
Finally we note that there may be upper limits to $\dot{M}$ or optical 
luminosity for producing [\ion{O}{3}] outflows. 
We plotted the lines which correspond to $\dot{M}=20 M_{\sun}$/yr and 
log $\lambda L_{\lambda} (5100 {\rm \AA}) ({\rm erg~ s}^{-1}) =45.8$ in Figure \ref{fig8}.
The small number of objects
with high Eddington ratios whose $\dot{M}$ or luminosity are larger than 
these lines suggests that these are not strong constraints.

\section{CONCLUSION} \label{Conclusion}
We have obtained optical spectra of the 
narrow-line quasars DMS 0059$-$0055 and PG 1543+489 with an intermediate
resolution ($R$=3000).
We discovered that DMS 0059$-$0055 has a large [\ion{O}{3}] blueshift 
relative to H$\beta$ (880 km s$^{-1}$), and
confirmed a large [\ion{O}{3}] blueshift in PG 1543+489 .
The [\ion{O}{3}] emission lines in both objects have $\sim 1000$ km s$^{-1}$
blueshifts relative to H$\beta$, and their widths are more than 1000 -- 2000 km s$^{-1}$.
These large blueshift and line width suggest that the outflow occurs in their nuclei
and interact with the ambient gas.
\par
We examined the correlations between $\Delta$v and the Eddington ratio as well as 
optical luminosity among ``blue outliers'', including
DMS 0059$-$0055 and other newly recognized ``blue outliers''.
There are no significant correlations of $\Delta$v with the Eddington ratio and optical luminosity
contrary to  naive expectations based on a radiative-pressure driven outflow hypothesis.
However, when we compared the Eddington ratio of ``blue outliers'' to those of other type 1 AGNs,
we confirm that ``blue outliers'' have the highest Eddington ratio among the 
objects with the same $M_{BH}$.
These facts suggest that the high Eddington ratio is a necessary condition 
or the $\Delta$v weakly depends on Eddington ratio
and the viewing angle may affect $\Delta$v.
In addition, we found that the higher $M_{BH}$ 
($> 10^{7} M_{\sun}$) and the higher mass accretion rate
($> 2 M_{\sun}$/yr) or the larger luminosity 
($\lambda L_{\lambda} (5100 {\rm \AA}) > 10^{44.6}$ erg s$^{-1}$) 
seem to be necessary to produce outflow besides high Eddington ratio.

\acknowledgments
We are grateful to the staffs of KPNO for their assistance during our observations.
We are also grateful to Anabela Gon\c{c}alves, Monique Joly, and Suzy Collin
for useful discussions and comments, and to Catherine Ishida for improving the manuscript.
T. K. is supported by the Japan Society for the Promotion of Science (JSPS)
Postdoctoral Fellowships for Research Abroad (464).
This research has made use of the NASA/IPAC Extragalactic Database (NED) 
which is operated by the Jet Propulsion
Laboratory, California Institute of Technology, under contract with the
National Aeronautics and Space Administration.
Funding for the creation and distribution of the SDSS Archive has been 
provided by the Alfred P. Sloan Foundation, 
the Participating Institutions, the National Aeronautics and Space Administration, the National Science Foundation, the U.S. Department of Energy, the 
Japanese Monbukagakusho, and the Max Planck Society. 
The SDSS Web site is \url{http://www.sdss.org/}. 
The SDSS is managed by the Astrophysical Research Consortium (ARC) for the Participating Institutions.
The Participating Institutions are The University of Chicago, Fermilab, the Institute for Advanced Study,
the Japan Participation Group, The Johns Hopkins University, Los Alamos National Laboratory, 
the Max-Planck-Institute for Astronomy (MPIA), the Max-Planck-Institute for Astrophysics (MPA), 
New Mexico State University, University of Pittsburgh, Princeton University, 
the United States Naval Observatory, and the University of Washington.

\clearpage

\clearpage 

\clearpage 

\clearpage 

\clearpage 

\clearpage 

\clearpage 

\clearpage 

\begin{deluxetable}{cccccccccc}
\tabletypesize{\scriptsize}
\rotate
\tablecaption{Blue Outliers \label{tbl1}}
\tablewidth{0pt}
\tablehead{
\colhead{Name} & \colhead{$\Delta$v} & \colhead{FWHM [O III]} &  
\colhead{FWHM H$\beta$} & \colhead{ref}  
& \colhead{log $\lambda L_{5100}$} & \colhead{log M$_{BH}$} & \colhead{log L$_{bol}$/L$_{Edd}$} 
& \colhead{log $P$} & \colhead{log $R$} \\
\colhead{} & \colhead{km s$^{-1}$}  & \colhead{km s$^{-1}$} & 
\colhead{km s$^{-1}$} & \colhead{} 
& \colhead{erg s$^{-1}$} & \colhead{M$_{\sun}$} & \colhead{} & \colhead{W Hz$^{-1}$}
&\colhead{} 
}
\startdata
PG 1402+261     & $-300$ & 900 & 1940 & 3 & 45.19 & 8.09 & 0.10 & 22.69 & $-0.67$ \\
IRAS 04416+1215 & $-300$\tablenotemark{a} & 1320 & 1470 & 4 & 45.11 & 7.79 & 0.32 & 23.18 & $-0.072$ \\
IRAS 04576+0912 & $-300$\tablenotemark{a} & 1290 & 1210 & 4 & 43.68 & 6.62 & 0.06 & \nodata & \nodata \\
PG 0804+761     & $-305$ & 780 & 3300 & 3 & 44.97 & 8.40 & $-0.42$ & 22.80 & $-0.33$ \\
RX J2217.9$-$5941 & $-330$\tablenotemark{b} & 1140 & 1370 & 3 & 44.92 &  7.59  & 0.32 & \nodata & \nodata \\
IRAS 13224$-$3809 & $-370$\tablenotemark{c} & 810 & $ > 650$\tablenotemark{d}  &1,2 
& 45.14 & 7.10 &  1.04 & 22.58 & $-0.70$ \\
RX J0136.9$-$3510 & $-380$ & 900 & 1050 & 3 & 44.73 &   7.24  & 0.50 & \nodata & \nodata \\
MS 2340.9$-$1511  & $-420$ & 780 & 970 & 3 & 44.50 &   7.01  & 0.50 & $< 22.68$ & $< 0.0070$ \\
PKS 0736+01     & $-430$ & 720 & 3260 & 3 & 44.98 &   8.39  & $-0.41$ & 26.23 & 3.1 \\
RX J2340.6$-$5329 & $-490$ & 780 & 1230 & 3 & 44.99 &   7.55  & 0.44 & \nodata & \nodata \\
RX J0439.7$-$4540 & $-580$ & 1020 & 1020 & 3 & 45.04 &   7.43  & 0.62 & \nodata & \nodata \\
PG 1415+451     & $-600$ & 660 & 2560 & 3 & 44.96 &   8.17  & $-0.21$ & 22.15 & $-0.97$ \\
I Zw 1          & $-640$ & 1440 & 1090 & 3 & 44.83 &   7.34  & 0.49  & 22.38 & $-0.59$ \\
Ton 28          & $-680$ &  960 & 1760 & 3 & 45.60 &   8.29  & 0.31 & \nodata & \nodata \\
DMS 0059$-$0055   & $-880$ & 1320 & 1500 & 5 & 45.28 &   7.93  & 0.35 & \nodata & \nodata \\
PG 1543+489     &$-1150$ & 2200 & 1630 & 5 & 45.59 &   8.05  & 0.37 & 23.86 & 0.0054 \\
\enddata
\tablenotetext{a}{These values were measured by us from the figure in \citet{VVG01}.}
\tablenotetext{b}{$-570$ km/s in \citet{gru01}}
\tablenotetext{c}{This value was measured by us from the figure in \citet{gru02}.}
\tablenotetext{d}{The H$\beta$ seems to be contaminated by H II region.
This value may be lower limit (K. Leighly, private communication).}
\tablerefs{
(1) Grupe \& Leighly 2002; (2) Leighly 1999b; (3) Marziani et al. 2003a; 
(4) V\'{e}ron-Cetty, V\'{e}ron, \& Gon\c{c}lves 2001; (5) this study 
}
\end{deluxetable}

\clearpage
\begin{deluxetable}{cccc}
\tabletypesize{\scriptsize}
\tablewidth{0pt}
\tablecolumns{4}
\tablecaption{Spearman's Rank Correlation Coefficients \label{tbl2}}
\tablehead{
\colhead{} & \colhead{} &
\colhead{$\Delta$v} &
\colhead{FWHM [O III]}
}
\startdata
$\Delta$v & $r_{s}$ & \nodata & $-0.18$ \nl
          & P$_{\rm r}$ & \nodata & 0.50 \nl
log L$_{bol}$/L$_{Edd}$ & $r_{s}$     & $-0.24$ & 0.25 \nl
                        & P$_{\rm r}$ &  0.37 & 0.35 \nl
log $\lambda L_{5100}$  & $r_{s}$     & $-0.37$ & 0.23 \nl
                        & P$_{\rm r}$ &  0.16 & 0.39 \nl
log $M_{BH}$ & $r_{s}$     & $-0.25$ & $-0.17$  \nl
             & P$_{\rm r}$ &  0.34 &  0.53  \nl
log $P$   & $r_{s}$     &  0.24 & 0.18   \nl
          & P$_{\rm r}$ &  0.57 & 0.64   \nl
log $R$   & $r_{s}$     & $-0.11$ & 0.075  \nl
          & P$_{\rm r}$ &  0.80 & 0.85   \nl
\enddata
\end{deluxetable}
\end{document}